\documentclass[a4paper]{llncs}

\usepackage{graphicx}
\usepackage{vdmlisting}
\usepackage{listing}
\usepackage{hyperref}
\usepackage{times}
\usepackage{url}
\usepackage{tcolorbox}
\usepackage{tipa}
\usepackage{gensymb}
\usepackage{enumitem}
\usepackage{xspace}

\newcommand{\remobidyc}{\textsc{re\textlengthmark mobidyc}\xspace}
\newcommand{\Remobidyc}{\textsc{Re\textlengthmark mobidyc}\xspace}
\newcommand{\ct}[1]{\textsf{#1}\xspace}

\lstdefinestyle{overtureLanguageStyle}{basicstyle=\footnotesize\ttfamily,
                        frame=trBL, 
                        tabsize=2, 
                        linewidth=\textwidth,
                        showstringspaces=false, 
                        captionpos=b,
                        frameround=fttt, 
                        aboveskip=5mm,
                        belowskip=5mm,
                        framexleftmargin=0mm, 
                        framexrightmargin=0mm,
                        escapeinside={(*@}{@*)},
                        language=VDM_SL}

\lstdefinelanguage{html}{ 
  backgroundcolor={\color[gray]{1}},
  basicstyle=\small\ttfamily,
  morekeywords={code}, 
  sensitive=true, 
  morestring=[s]{"}{"}, 
  style=HtmlStyle 
} 
\lstdefinestyle{HtmlStyle}{ 
} 
\lstdefinelanguage{Smalltalk}{ 
  basicstyle=\small\ttfamily,
  morekeywords={true,false,self,super,nil}, 
  sensitive=true, 
  morecomment=[s]{"}{"}, 
  morestring=[d]', 
  style=SmalltalkStyle 
} 
\lstdefinestyle{SmalltalkStyle}{ 
} 

\lstdefinelanguage{remobidyc}{ 
  basicstyle=\normalsize\ttfamily,
  otherkeywords={`,=,->},
  morekeywords={to,is,with,where,when,the,uniform,normal, sigma,gamma,mean,scale,rate}, 
  sensitive=true, 
  style=remobidycStyle 
} 
\lstdefinestyle{remobidycStyle}{ 
  literate={degreeC}{\degree{}C}2 {degreeF}{\degree{}F}2 {delta}{$\Delta$}1,
}

\title{Implementation-First Approach of Developing Formal Semantics of a Simulation Language in VDM-SL}

\titlerunning{21th Overture Workshop, 2021}

\author{
Tomohiro Oda\inst{1} \and 
Ga{\"e}l Dur\inst{2} \and
St{\'e}phane Ducasse\inst{3} \and
Hugo Daniel Macedo\inst{4}
}

\authorrunning{Oda, T.; Dur, G.; Ducasse, S.; Macedo, H. M.}

\institute{Software Research Associates, Inc.~(\email{tomohiro@sra.co.jp})
\and Creative Science Unit (Geoscience), Faculty of Science, Shizuoka University~(\email{dur.gael@shizuoka.ac.jp})
\and Univ. Lille, Inria, CNRS, Centrale Lille, UMR 9189 -- CRIStAL, France~(\email{stephane.ducasse@inria.fr})
\and Aarhus University, DIGIT, Department of Electrical and Computer Engineering,~(\email{hdm@ece.au.dk})
}
\begin{document}

\maketitle

\begin{abstract}
Formal specification is a basis for rigorous software implementation.
VDM-SL is a formal specification language with an extensive executable subset.
Successful cases of VDM-family including VDM-SL have shown that producing a well-tested executable specification can reduce the cost of the implementation phase.
This paper introduces and discusses the reversed order of specification and implementation.
The development of a multi-agent simulation language called \remobidyc is described and examined as a case study of defining a formal specification after initial implementation and reflecting the specification into the implementation code.
\end{abstract}

%%%%%%%%%%%%%%%%%%%%%%%%%%%%%%%%%%
\section{Introduction}
\label{sec:Introduction}

Lightweight formal methods  are partial applications of formal methods so that a formal specification of the whole or a part of the software system provides the implementation with clear goal conditions of its functionality~\cite{Agerholm&98d,Fitzgerald&08a}.
A formal specification consists of concise and unambiguous definitions of the system's functions which allow implementors to focus on the correctness and efficiency of the implementation.
VDM-SL~\cite{Larsen&13b} is one of the oldest formal specification languages still in practical use.
VDM-SL has an extensive subset that can be executed by interpreters, which enables unit testing of specifications.
Successful cases of the VDM-family~\cite{Fitzgerald&07e,Oda&18} have shown that producing a well-tested executable specification benefits the implementation phase.
Unit testing ensures that the algorithms defined in the specification work as expected in the test cases.
Test cases for the VDM specification can also be rewritten in implementation languages so that the implementation works the same as the executable specification.

While the specification is deemed to be written before the implementation in traditional views of software lifecycles, the implementation sometimes precedes the specification in practice.
Development of a GUI application is sometimes driven by GUI prototypes before the developers capture the whole required functionality of the application.
This then raises the following questions: 
Is it worth writing a formal specification even if the implementation is already running?
How much does it cost to write a concise specification based on an exploratory implementation?

This paper describes a development process where the formal specification was written after the implementation.
\Remobidyc~\cite{oda2021re} is a multi-agent simulation system for population dynamics in biology.
Defining the operational semantics of the modeling language in VDM-SL was planned from the beginning of the development.
Although the implementation phase typically follows the formal specification phase,  the implementation of a GUI-based environment including an interpreter was started without the formal semantics.
Domain-specific features in the modeling languages, such as life-history stages, moulting and reproduction, were implemented and evaluated in exploratory manners involving biologists.
The formal semantics was defined after the language features had become stable.
The formal specification was refactored into a more concise presentation and the implementation was improved by reflecting the refactored specification.

In Section~\ref{sec:remobidyc}, the design and implementation of \remobidyc and its modeling language are explained.
Its development process will be described in Section~\ref{sec:development-process}.
Findings from the development will be discussed in Section \ref{sec:discussions}, and Section \ref{sec:concluding-remarks} concludes the paper.

%%%%%%%%%%%%%%%%%%%%%%%%%%%%%%%%%%
\section{\Remobidyc and its modeling Language}
\label{sec:remobidyc}

\Remobidyc is a multi-agent simulation platform for the scientific study of biology and ecology, built upon Pharo~\cite{oda2021re}.
In multi-agent simulation, agents with simpler definitions interact with each other to exhibit complex phenomena.

\begin{figure}
\begin{center}
\includegraphics[width=1\textwidth]{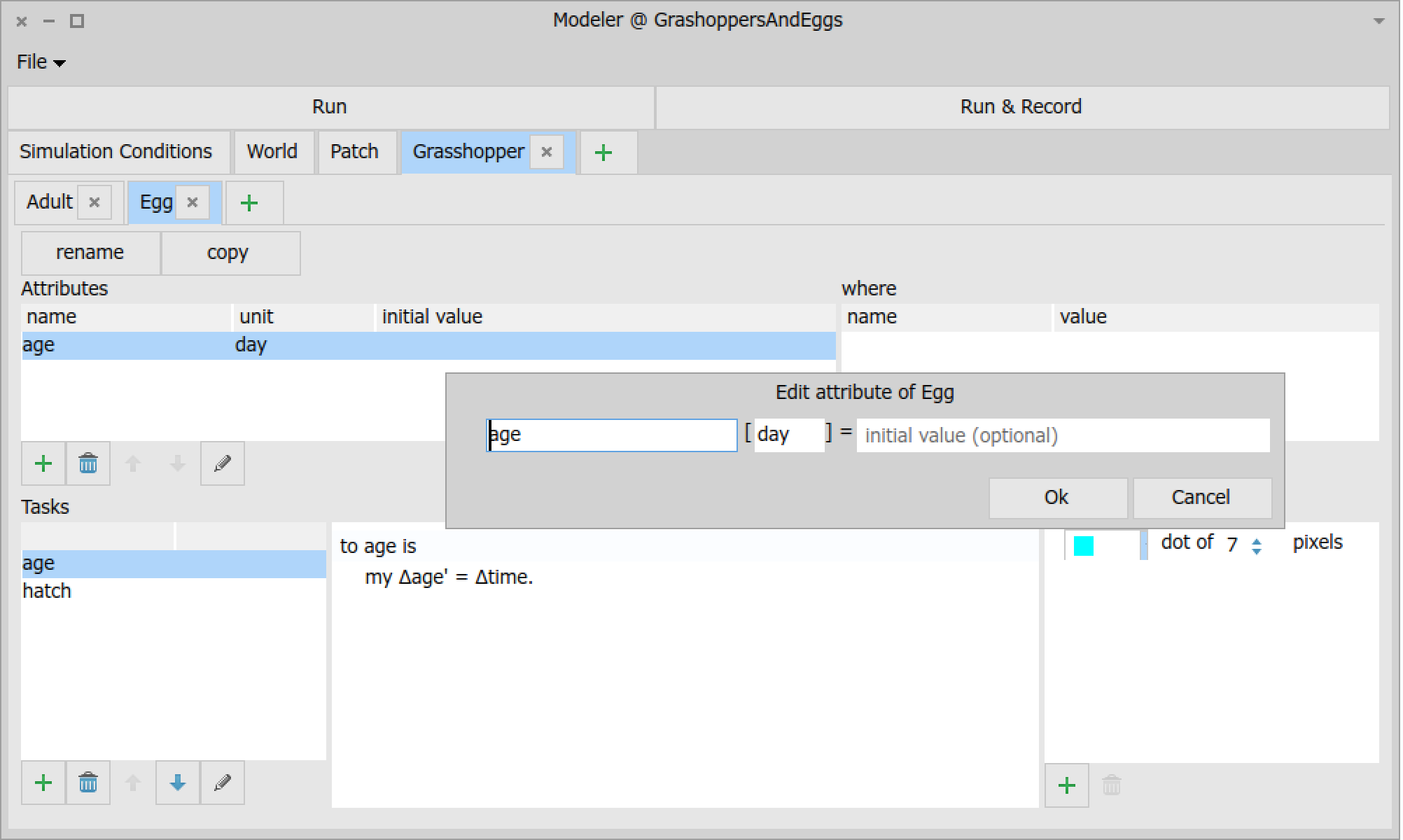}
\end{center}
\caption{Screenshot of \remobidyc's GUI-based modeler}
\label{fig:screenshot-gui-modeler}
\end{figure}

The following are \remobidyc's design principles to support the user's tasks.
\begin{enumerate}[leftmargin=3em,label=DP\arabic*.]
\item The language should not impose imperative programming.\label{enum:math-friendly}
\item The language must guarantee the computation always terminates or aborts.\label{enum:no-loops}
\item The system should provide GUI-based interfaces to define a model. \label{enum:gui-based-modeler}
\item The language should provide semantic checking based on measurement units.\label{enum:checking}
\item The system should record all attribute variables of all agents at all time steps.\label{enum:record-all-attributes}
\item The system should always produce the same result from the same model.\label{enum:reproducibility}
\end{enumerate}
DP1 and DP2 are to support biologists and ecologists who are not necessarily familiar with imperative programming languages.
In general, it is hard to verify whether or not an imperative program with assignments and conditional loops implements the mathematical model at hand.
The termination property is also hard to verify.
DP2 implies that the language will not be Turing complete.
In return, the model will be free of incidental infinite loops.
DP3 and DP4 help the user to reduce the burdens of syntax and semantic errors in the model.
Fig. \ref{fig:screenshot-gui-modeler} shows the GUI-based modeler for DP3.
The basic idea of DP4 is to use measurement units as static types to detect semantically ill-formed expressions in the model.
DP5 and DP6 are for the verifiability and reproducibility of simulation results to make the simulation more useful as a tool for scientific research.

\begin{figure}
\begin{center}
\includegraphics[width=1\textwidth]{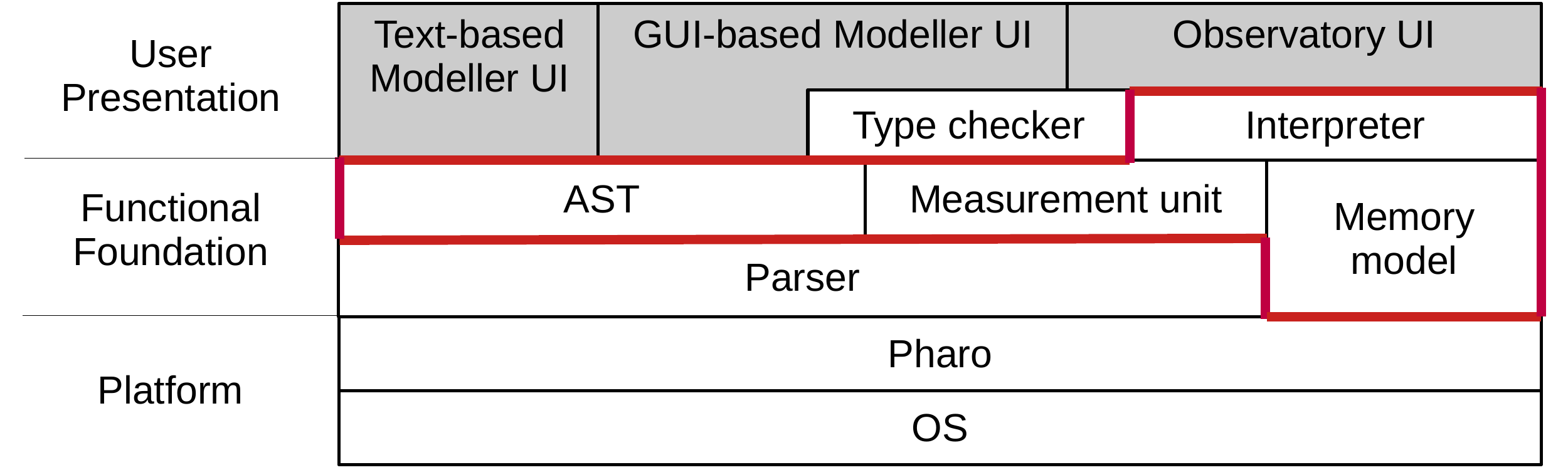}
\end{center}
\caption{Major components of \remobidyc simulation environment}
\label{fig:major-components}
\end{figure}

\Remobidyc is built upon Pharo~\cite{Berg13a,Blac09a}, a modern dynamic object-oriented language derived from Smalltalk, which provides a flexible and immersive programming environment on macOS, Linux and Windows systems.
Fig. \ref{fig:major-components} illustrates the software configuration of the \remobidyc simulation environment.
It shows three major UI elements, namely text-based modeler, GUI-based modeler, and Observatory which runs a simulation and visualises the result.
Functional components operated by the user such as type checker and interpreter are also built on the fundamental elements such as the abstract syntax tree (AST), measurement units and the memory model.
Among the UIs, text-based modeler is provided as a backup to GUI-based modeler UI in the case that the model definition files are broken for some reason.
The scope of the formal specification in VDM-SL was limited to the components enclosed by the red lines in Fig. \ref{fig:major-components}.
The specification phase was started after their first implementation in Pharo had been done and the language had become stable enough.

The development of \remobidyc required a GUI-based modeling environment and exploratory design of the modeling language because the language must be validated by experts from biology and ecology.
Pharo's dynamic nature enables agile prototyping to quickly develop the GUI-based modeler and the interpreter.
The modeling language and its interpreter including the memory model were developed with the implementation in Pharo first, and then its formal semantics was defined in VDM-SL.
The development is still ongoing in both implementation and formal semantics as an open-source project.

%%%%%%%%%%%%%%%%
\subsection{Overview of the Language Design}
\Remobidyc is designed to be friendly to mathematical modelers in biology and ecology, demanding less programming skills\cite{oda2021re}.
The models in \remobidyc are constructed in the declarative manner so that the models look more like definitions of numeric sequence in the form of recurrence relations rather than a series of imperative statements.

\begin{figure}
\begin{center}
\includegraphics[width=0.65\textwidth]{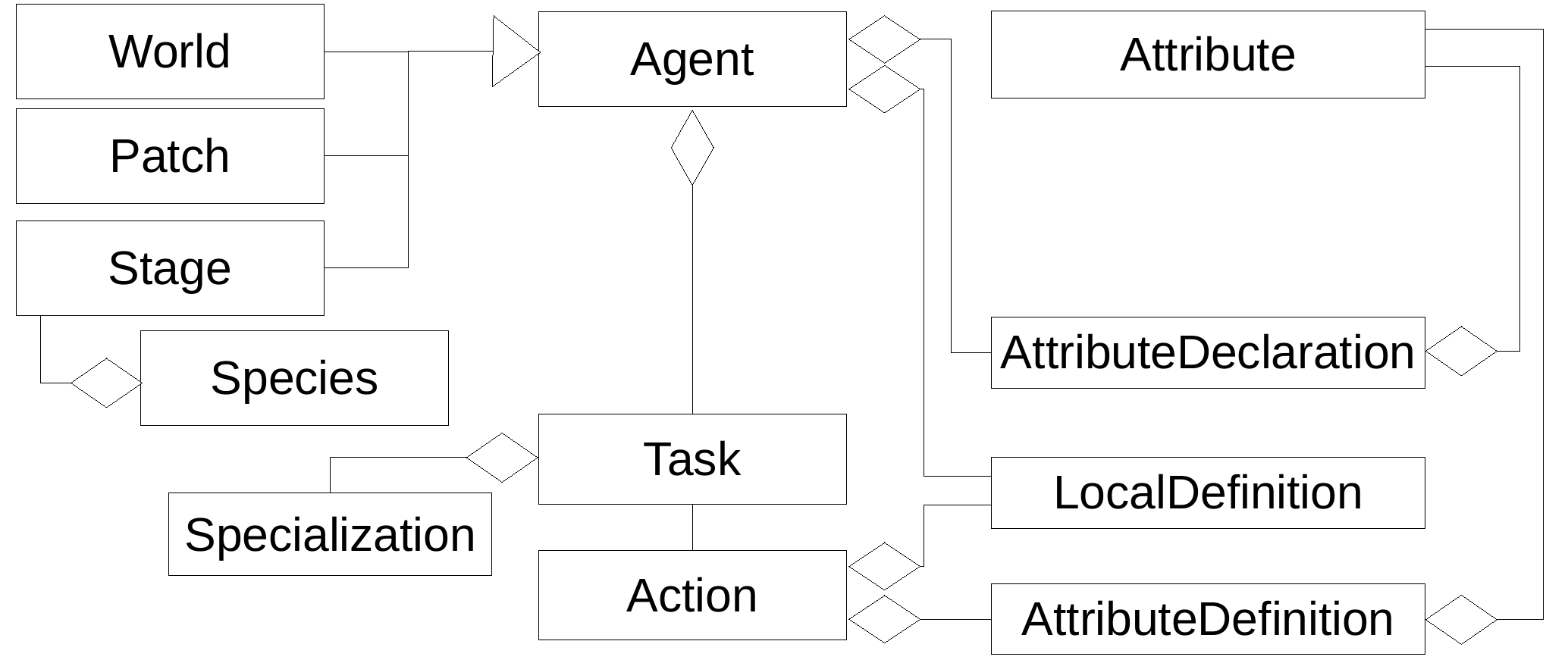}
\end{center}
\caption{Class diagram of re:mobidyc's modeling language AST}
\label{fig:class-diagram}
\end{figure}

Fig. \ref{fig:class-diagram} illustrates the major constructs of the modeling language.
Agent, action, task, and attribute are the most significant constructs in the modeling language.
\Remobidyc has three kinds of agents: \ct{World}, \ct{Patch}, and \ct{Stage}.
\ct{World} is the agent that represents the global environment, and a \ct{Patch} represents the local environment.
A {\it stage} is also called an {\it animat}, which has two attributes \ct{x} and \ct{y} so that they can move around in a two-dimensional space.
A stage represents a life-history stage of an individual life such as larva, juvenile, and adult which together represent an individual.
An individual is modeled as an instance of a {\it species}.

Each agent has a set of {\it attribute declarations} each of which specifies the identifier, the measurement unit and optionally the initial value of the \ct{attribute} allocated in the persistent heap memory.
An interaction among agents is modeled as an {\it Action} which consists of a set of {\it attribute definitions}, each of which modifies the next value of the attribute of an animat that participates in the interaction.
Local variables are defined in the {\it utility definitions} so that right-hand expressions in attribute declarations and attribute definitions are concisely presented.
A \ct{task} definition binds an action to an agent so that each individual of the agent will do the action at every time step in the simulation.

\begin{figure}
\begin{center}
\includegraphics[width=1\textwidth]{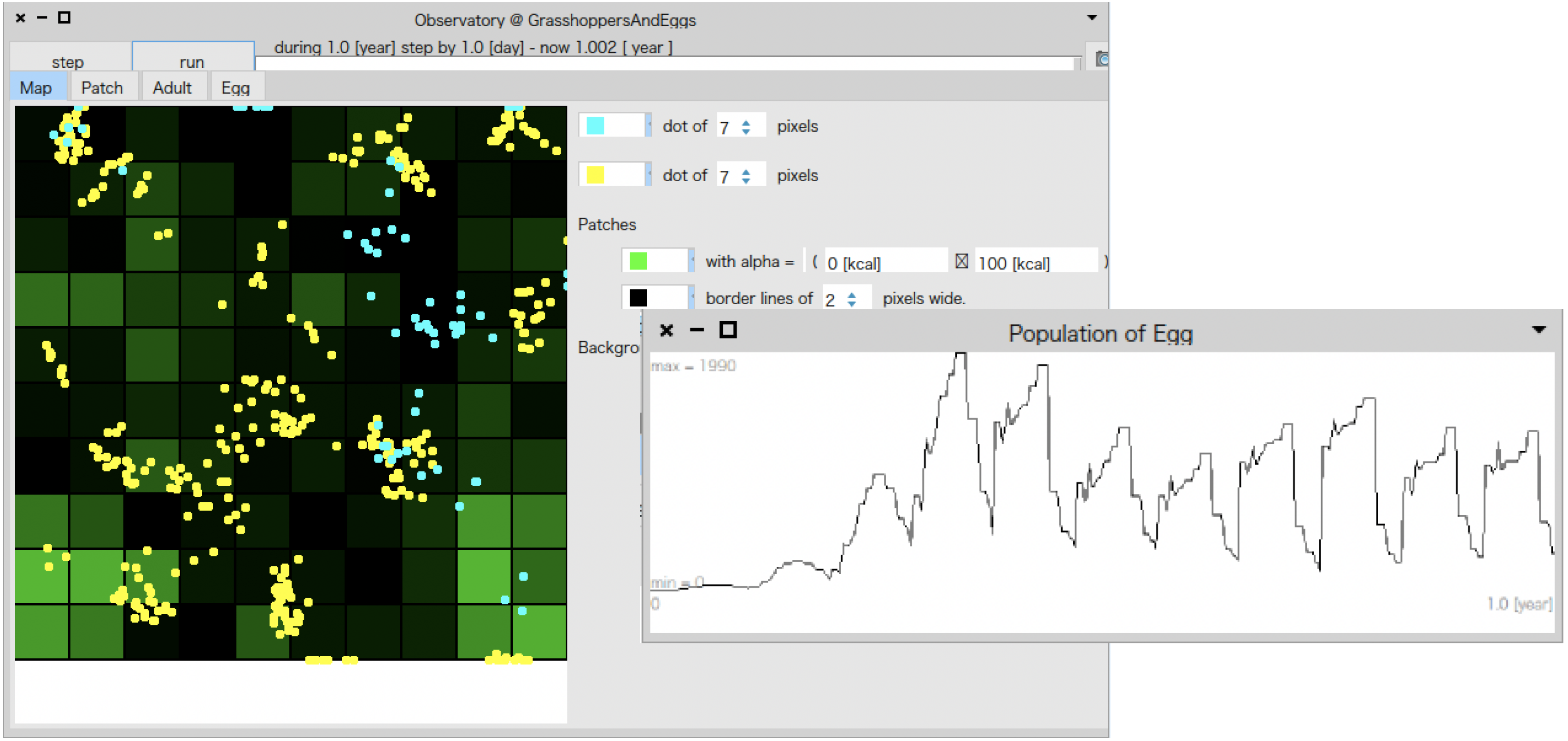}
\end{center}
\caption{Screenshot of re:mobidyc's Observatory}
\label{fig:screenshot-observatory}
\end{figure}

Fig. \ref{fig:screenshot-observatory} shows a screenshot of \remobidyc running ``eggs and grasshoppers'', a simple model to explain the basic features of \remobidyc.
The space is divided into rectangular patches, visualized as green rectangles in the figure, each of which has grass on it.
A grasshopper, rendered as yellow dots in the figure, moves to the cell with the richest grass and eats it.
Eating grass reduces the grass in the patch.
The grasshopper also stores energy from food and metabolism reduces the energy at a certain rate.
If the energy is below a threshold, the grasshopper starves to death.
The grasshopper spawns eggs when it gets matured.
An egg, rendered as cyan dots in the figure, does not do anything until it eventually hatches after a certain duration, and becomes a grasshopper.
A chart of the population of eggs by time is displayed to the right.
Details of these language constructs are explained later.

%%%%%%%%%%%%%%%%
\subsection{Agents and Attributes}

An agent has a set of attributes, each of which consists of an identifier, a measurement unit, and an optional initial value.
The measurement unit is used for type checking to be described in Section~\ref{sec:type-system} and also to display numbers in visualization such as tables and charts.
An attribute is a variable that stores a floating point number.
All numbers in attributes and expressions of \remobidyc are in the SI unit.
Stages implicitly have the \ct{x} and \ct{y} attributes in \ct{m} to hold the position of the agent.
\ct{World} represents the global environment and \ct{Patch} represents a local environment of a square region of the space.

\begin{figure}
\begin{lstlisting}[language=remobidyc]
Adult is Grasshopper with
    age [day].
Egg is Grasshopper with
    age [day] = 0 [day].
\end{lstlisting}
\caption{Example definitions of two stages in \remobidyc}
\label{fig:example-animat}
\end{figure}

Fig. \ref{fig:example-animat} shows the concrete syntax of the agent definitions  in the eggs and grasshoppers model.
The first and the second lines define the \ct{Adult} stage of the \ct{Grasshopper} species with the \ct{age} attribute in \ct{day}.
The third and the fourth lines define the \ct{Egg} stage of the \ct{Grasshopper} species also with the \ct{age} attribute in \ct{day} initialized with \ct{0 [day]}.

%%%%%%%%%%%%%%%%
\subsection{Actions and Tasks}
\label{sec:actions-and-tasks}

Actions are core constructs of a model that represent interactions among agents as modifications to their attribute variables.
\Remobidyc employs discrete-event simulation with {\it synchronous} updates on memory, which holds all modifications to attributes during the computation at a time step.

To implement the synchronous updating memory, \remobidyc allocates three memory slots for an attribute: the \textit{value} slot, the \textit{next} slot, and the \textit{delta} slot.
\begin{itemize}
\item The value slot is for the read access.
\item The next slot and the next-delta slot are for the write access.
\end{itemize}

The interpreter can overwrite the next slot with a new value of the attribute for the next time step.
The value will not be read until the next time step.
The delta slot accumulates the increments/decrements to the attribute.
The increments/decrements to an attribute across different actions will not affect the attribute until the next time step but are accumulated in the delta slot.
At the end of the time step, the value slot is updated with the sum of the next slot and the delta slot.
The details of the synchronous updating memory are explained in Section~\ref{sec:memory-model}.

The synchronous updating memory was chosen due to DP5: {\it The system should record all attribute variables of all agents at all time steps}.
The record of attribute variables at every time step carries all information about the simulation because the values of the attribute variables can change only at the transitions between time steps.
Also, the synchronous updating semantics makes the semantics of actions closer to definitions of numeric sequence in the form of recurrence relation because neither the execution order of actions nor the order of agents affects the result of the simulation.
All changes are accumulated and reflected regardless of the order of execution.

\begin{figure}
\begin{lstlisting}[language=remobidyc]
to age is
    my delta age' = delta time.
to move is
    my d/dt x' = cos(theta)*r
    my d/dt y' = sin(theta)*r
where
    theta = the heading
    r = the speed.
\end{lstlisting}
\caption{An example definition of actions in \remobidyc}
\label{fig:example-action}
\end{figure}

Fig. \ref{fig:example-action} shows the definitions of two actions, namely \ct{age} and \ct{move} in the \remobidyc modeling language.
The \ct{age} action has one attribute definition to add \ct{$\Delta$time}, the simulation time step defined as a simulation setting, to the \ct{age} attribute of the agent that performs the action.
Please note that which agent will perform this action is not specified yet.
An agent and an action are bound by a task definition, which is described later in this section.

The \ct{move} action modifies the attributes \ct{x} and \ct{y} of the performing agent using two utility variables \ct{theta} and \ct{r} defined below the \ct{where} clause.
A utility variable is a kind of temporary variable scoped within the action.
Identifiers led by \ct{the} prefix, such as \ct{the heading} and \ct{the speed}, are placeholders which will be replaced with concrete expressions in a task definition.
Placeholder allows an action to have a generic parameter whose concrete argument may vary by the performer agent.
Above the \ct{where} clause, two attribute definitions on the \ct{x} and \ct{y} attributes are placed.
An attribute definition with the \ct{d/dt} decorator is equivalent to the attribute definition with \ct{$\Delta$} operator and the right-hand side multiplied by the simulation time step \ct{$\Delta$time}.
For example, \ct{my d/dt x$'$ = cos(theta)*r} is equivalent to \ct{my $\Delta$ x$'$ = cos(theta)*r*$\Delta$time}.

\begin{figure}
\begin{lstlisting}[language=remobidyc]
Adult move
where
    the speed -> uniform 0 [km/day] to 0.5 [km/day]
    the heading -> direction neighbor's grass.
\end{lstlisting}
\caption{An example definition of a task in \remobidyc}
\label{fig:example-task}
\end{figure}

As stated in DP1: {\it The language should be declarative and should not impose imperative programming}, the action definition is in a declarative style.
Assignments are listed at the top level of an action definition, and the right-hand sides are expressions which do not contain assignments.
Built-in functions except random number generators in uniform, normal, gamma and log-logistic distributions are referentially transparent, and the interpreter manages internal states of random number generators to ensure reproducibility.
Expressions such as arithmetics and built-in functions trivially correspond to their counterparts in mathematics.
All values are floating point numbers and neither booleans nor agents are the first-class objects.
No loop can appear in expressions including finding the local patch or nearby agents so that an execution of a simulation should halt or abort in a finite time.
Finding a peer for interaction is done by the interpreter.

%%%%%%%%%%%%%%%%
\subsection{Formal Semantics of the Language}

Although the language is intended to be easy to learn without prerequisite programming skills, its semantics should be defined without ambiguity as a tool for scientific research.
For example, the internal state of the random number generator depends on the number of random numbers generated so far even when the same seed is specified.
A precise process of execution must be specified to clarify, for example, when the right-hand side of a utility definition is evaluated and how many times.

\begin{figure}
\begin{vdmsl}
types
  AttributeDefinition ::
    variable : AttributeVariable | Placeholder
    decorator : Decorator
    expression : Expression;
  AttributeVariable :: 
     agent : [Identifier] 
     identifier : Identifier;
  Decorator = <assign> | <delta> | <differential>;
  Expression =
    Variable | Literal | Casting | Apply | Arithmetics | ...;
\end{vdmsl}
\caption{AST definition of actions in VDM-SL}
\label{fig:actions-vdm}
\end{figure}

\paragraph{Syntactical definitions.}
Fig. \ref{fig:actions-vdm} shows the definition of attribute definition's AST, and Fig. \ref{fig:eval-vdm} defines how variable references in expressions are evaluated.
Triple dots in the VDM-SL specification indicate omission.
The main objectives of defining the formal semantics of \remobidyc are not in mathematical proof of certain properties of the language, but to provide a clear reference of the language for understanding models and for ensuring compatibility of ported interpreters in future.

\begin{figure}
\begin{vdmsl}
  evalExpression : AST`Expression ==> real
  evalExpression(expr) ==
    cases expr:
      mk_AST`UtilityVariable(identifier) ->
        let val = Interpreter`readUtility(identifier)
        in
          (if val <> nil then return val;
          let newval = evalExpression(
            Interpreter`getUtilityDefinition(identifier))
          in 
            (Interpreter`writeUtility(identifier, newval);
            return newval)),
      mk_AST`AttributeVariable(agent, identifier) ->
        return Memory`read(
          Interpreter`getAttributeAddress(agent, identifier)),
      ...
    end;
    
  readVariable : [AST`Identifier] * AST`Identifier ==> real
  readVariable(agent, identifier) ==
    return Memory`read(
      Interpreter`getAttributeAddress(agent, identifier));
    
  evalAttributeDefinition : AST`AttributeDefinition ==> ()
  evalAttributeDefinition(attributeDefinition) ==
    let
      value = evalExpression(attributeDefinition.expression),
      agent = attributeDefinition.variable.agent,
      identifier = attributeDefinition.variable.identifier
    in
      cases attributeDefinition.decorator:
        <assign> -> writeVariable(agent, identifier, value),
        <delta> -> writeDeltaVariable(agent, identifier, value),
        <differential> -> 
          writeDeltaVariable(agent,identifier,value*deltaTime())
      end;
                
  writeVariable:[AST`Identifier]*AST`Identifier*real==>()
  writeVariable(agent, identifier, data) ==
    Memory`write(
      Interpreter`getAttributeAddress(agent, identifier), 
      data);
    
  writeDeltaVariable:[AST`Identifier]*AST`Identifier*real==>()
  writeDeltaVariable(agent, identifier, data) ==
    Memory`writeDelta(
      Interpreter`getAttributeAddress(agent, identifier), 
      data);
\end{vdmsl}
\caption{Semantics of variable references in VDM-SL}
\label{fig:eval-vdm}
\end{figure}

\paragraph{Execution semantics.}

The definitions of major operations related to evaluating expressions and attribute definitions are shown in Fig. \ref{fig:eval-vdm}.
\Remobidyc provides expressions such as variable references, literals, built-in functions, arithmetic operators and so on.
The definition body of the \ct{evalExpression} operation is a huge cases statement that defines how to evaluate each kind of expression.
The semantics of variables with regard to read and write access is a core feature of \remobidyc.

The reference to a utility variable is defined in three steps: (1) try to read first and return if successful, (2) evaluate the right-hand side of the utility definition, and (3) store the result and return it.
Thus, utility variables are evaluated on demand and written once for each evaluation of the action.
The reference to an attribute variable defined by the \ct{readVariable} operation states that the interpreter is also responsible for resolving the address of the attribute variable of an agent.
It is also clear that the values of the utility variable are managed by the \ct{Interpreter} module while the attribute variables are allocated in the \ct{Memory} module.

The definitions of the \ct{writeVariable} and \ct{writeDeltaVariable} operations define the write access to the \ct{next} slot and the \ct{delta} slot accordingly, and the interpreter resolves the address of the specified attribute of the agent.
The definition of the \ct{evalAttributeDefinition} operation specifies the mapping between the decorators and memory's write-access API.

%%%%%%%%%%%%%%%%
\subsection{Type System}
\label{sec:type-system}

The modeling language of \remobidyc has a unique type system based on measurement units.
All attribute variables, utility variables, and expressions in \remobidyc are floating point numbers with measurement units.
All values are computed in SI units, and a pair of types are compatible when their measurement units have the same dimension.
For example, an expression \ct{10 [km] / 3 [h]} is typed \ct{[km/h]} with dimension \ct{[m/s]}.
\ct{10 [km] + 3 [h]} is a type error because the \ct{+} operator requires the both arguments in compatible types and \ct{[km]} and \ct{[h]} are not compatible.
Two kinds of type casting expressions are provided; the en-unit conversion to attach a measurement unit to a non-dimensional number typed \ct{[]}, and the de-unit conversion to detach the measurement unit to generate a non-dimensional number.
These type castings are useful when dealing with expressions with exponentials and logarithms.

\begin{figure}
\begin{vdmsl}
types
  Unit :: dimension:seq of (SIBaseUnit * int) scale : real
    inv mk_Unit(us, -) ==
      card {u | mk_(u, -) in seq us} = len us
      and (forall mk_(-, o) in set elems us & o <> 0);
  SIBaseUnit = <kg>|<m>|<s>|<degreeC>|<K>|<degreeF>|<rad>|<mol>;
\end{vdmsl}
\caption{The definition of measurement units in VDM-SL}
\label{fig:measurement-units-vdm}
\end{figure}

Fig. \ref{fig:measurement-units-vdm} shows a snippet from the definition of measurement units in VDM-SL.
The \ct{Unit} type has two fields: the dimension field which has a sequence of SI base units and their orders, and the scale field for unit conversions in the type casting expressions and literal values with non-SI units.
For example, the unit of speed \ct{[km/h]} is represented as \ct{mk\_Unit([mk\_(<m>, 1), mk\_(<s>, -1)], 0.27777...)}.
Although the temperature unit \ct{<degreeC>} is not an SI base unit, \remobidyc handles it as if it were an SI base unit because the Celsius temperature is widely used in math models of Biology, e.g. {\it degree day}. The same applies to the Fahrenheit temperature.

\begin{figure}
\begin{vdmsl}
types
     Address = nat1;

state Memory of
    vals : map Address to real
    next : map Address to real
    delta : map Address to real
    ...
    valuesStorage : seq of (map Address to real)
    animatsStorage : seq of (map Address to (AST`Identifier * nat1))
    ticks : nat
init s == ...
end

operations
  pure read : Address ==> real
  read(address) ==
    if address in set dom vals 
    then return vals(address) 
    else exit ADDRESS_ERROR;
    
  write : Address * real ==> ()
  write(address, data) == next(address) := data;
    
  writeDelta : Address * real ==> ()
  writeDelta(address, data) ==
    if address in set dom delta
    then delta(address) := delta(address) + data
    else exit ADDRESS_ERROR;

  store : () ==> ()
  store() ==
    (valuesStorage := valuesStorage
      ^[{a|->next(a)+(if a in set dom delta then delta(a) else 0)
        | a in set dom next \ deads}];
    ...)
  pre  ticks = len valuesStorage and ticks = len animatsStorage;

  load : nat1 ==> ()
  load(t) ==
    (vals := valuesStorage(t);
    ...
    next := vals;
    delta := {a |-> 0 | a in set dom vals};
    ...
    ticks := t)
  pre  t <= len valuesStorage and t <= len animatsStorage;
\end{vdmsl}
\caption{The definition of read and write access to the memory in VDM-SL}
\label{fig:memory-model-slots-vdm}
\end{figure}

%%%%%%%%%%%%%%%%
\subsection{Memory}
\label{sec:memory-model}

As explained in Section \ref{sec:actions-and-tasks}, \remobidyc employs synchronous updates on memory.
All modifications to attribute variables are delayed until the end of the time step.
The values of all attributes are recorded to storage as required by DP5: {\it the system should record all attribute variables of all agents at all time steps}.

Fig. \ref{fig:memory-model-slots-vdm} shows the specification of memory access in VDM-SL.
The memory holds three mappings from \ct{Address} to \ct{real} as the memory's internal state.
The state variables \ct{vals}, \ct{next} and \ct{delta} hold the value slots of values, next and delta accordingly.
The three operations \ct{read}, \ct{write} and \ct{writeDelta} define the three kinds of access to the memory.
Because operations are not first-class objects in VDM-SL, callers to the \ct{write} and \ct{writeDelta} operations can be statically analyzed to ensure the evaluation of expressions does not reach the \ct{write} and \ct{writeDelta} operations.

The values of all attributes are stored into storage.
The \ct{store} operation appends the sum of the next slot and the delta slot into the state variable \ct{valuesStorage} typed as a sequence of real numbers.
The \ct{load} operation initializes the values slots and the next slots with the stored values, and sets the delta slots to zero.
The \ct{store} and \ct{load} operations are called at transitions of simulation time steps to synchronize the three slots and the storage.
These two operations abstract the implementation of the storage \ct{valuesStorage}.
The \ct{load} operation takes the time step as the argument, which allows the simulation not only to proceed forward but also to be unwound to a point in the past to replay the simulation.
In Pharo, the Observatory UI shown in Fig. \ref{fig:screenshot-observatory} has a slider to control the simulation time step back and forth.

The formal specification of the memory brings the significant benefit of enabling multiple storage back-ends.
In Pharo, two kinds of storage are implemented.
One is on-memory storage which simply stores the values as an ordered collection object which runs fast but has the limitation of capacity.
Small models with short simulation time run efficiently on the on-memory storage.
Another is file-based storage which dumps the values into CSV files.
The file-based storage uses less memory regardless of the number of time steps in the simulation, at the cost of the reading and writing files.
It is also possible to implement a storage back-end on RDBMS shared by multiple users.
In Pharo, an abstract class for storage defines the public API of storage classes.
The formal specification in VDM-SL adds the semantic requirements of the API so that the user can safely choose any concrete storage.

%%%%%%%%%%%%%%%%%%%%%%%%%%%%%%%%%%
\section{Development Process}
\label{sec:development-process}

The development of \remobidyc is hosted on the github organization~\footnote[1]{\url{https://github.com/ReMobidyc/}}.
Pharo source code and VDM-SL specification are together in the same repository.

\begin{table}
\caption{Source size during development time}
\label{table:source-size}
\begin{tabular}{ l | c || r | r | r | r | r } 
event 		& date 		& Pharo LOC	& Pharo LOC & Pharo tests 	& VDM LOC	& VDM tests\\
			&			& (all)		& (interp.) 	& (interp.)	&			&\\
\hline
impl.  started 	& Oct 2019 	& - 			& -			& -			& - 			& -\\
			& Jan 2020	& 1,499		& 1,150		& 21			& - 			& -\\
			& Jan 2021	& 12,936 		& 7,268		& 214		& - 			& - \\
			& Jan 2022	& 19,474 		& 9,526		& 276		& - 			& -\\
spec started 	& Aug 2022 	& 26,330 	& 11,990		& 320		& 0 			& 0\\
 			& Dec 2022 	& 30,114 		& 13,205		& 338 		& 1,364 		& 113\\
\end{tabular}
\end{table}

Table~\ref{table:source-size} shows the progress of the development in the sizes of implementation and specification.
The {\it Pharo LOC (all)} column shows the total number of lines of source code stored in the repository, including GUIs, parsers and type checkers.
The {\it Pharo LOC (interp.)} column shows the number of lines of code for interpretation, {\it i.e.} AST, evaluation and memory model.
The {\it VDM LOC} column shows the number of lines of the VDM specification for the corresponding parts.

By reading the time factors, the specification went far quicker than the implementation.
Implementation in Pharo started in Oct 2019 and is still ongoing.
The specification in VDM-SL started in Aug 2022, about 3 years after the start of the implementation by the same engineer who implemented it in Pharo.
The specification was developed on ViennaTalk~\footnote[2]{\url{https://github.com/tomooda/ViennaTalk}}~\cite{Oda&17a} and was also refined on the Overture VSCode plugin~\footnote[3]{\url{https://github.com/overturetool/vdm-vscode}}~\cite{Rask&20,Rask&22}.
The implementation for interpretation gradually increased throughout the development time and the specification caught up in four months.
Please note that this should not be taken as the difference in productivity between Pharo and VDM.
The difference reflects the cost of exploration to discover the language features validated by the domain experts.
Building GUIs, designing concrete syntax, writing parsers, implementing various visualizations, creating type-checking algorithms, and communicating with domain experts are involved in the exploratory process.

The LOCs also indicate that the code in Pharo is about 10 times larger than its counterpart in VDM-SL.
This also does not mean that the VDM is 10 times more productive than Pharo.
Both Pharo and VDM provide high-level functions to manipulate abstract concepts such as finite sets and mappings.
One source of the difference is the redundancy of Pharo's file format.
Editing source files is not a typical way of programming in Pharo.
Pharo provides powerful programming UIs that the programmer can interactively write code using automated tools.
Accessor methods to read and write instance variables can be automatically generated by the tool, which takes a significant amount of lines when stored in a file.

\section{Discussion}
\label{sec:discussions}
Successful cases of formal methods including VDM have shown that formal specification benefits the implementation phase in time, cost and quality~\cite{Fitzgerald&07e}.
The development of \remobidyc went in the reverse order of the specification and its implementation.
The specification of \remobidyc plays two roles in the development: an intermediate product as input to the implementation, and the specification itself is a part of the final product.
The language is designed as a tool for scientific research and therefore its semantics should be rigorously defined and available to the open public.

Writing the specification from implementation went smoothly despite the difference in paradigms between Pharo and VDM-SL.
Pharo is a dynamically-typed object-oriented language while VDM-SL is modularized definitions of types, values, functions, states, and operations without classes.
There were minor mismatches in the way of the presentation of concepts between Pharo and VDM-SL.
One is {\it kind-of} relations presented in a class hierarchy in Pharo.
In VDM, they are presented as subtypes using union types.
For example, an agent is either the world, a patch or a stage in \remobidyc.
In Pharo, the classes for world definition, patch definition and stage definition are subclasses of the agent definition class.
In VDM-SL, the \ct{AgentDefinition} type is the union of \ct{WorldDefinition}, \ct{PatchDefinition} and \ct{AnimatDefinition}.
These subtype relations implemented in the class hierarchy are naturally rewritten as union types in VDM-SL.

An apparent difference in the presentation of implementation and specification is in how functions on AST nodes are defined.
Although both Pharo and VDM-SL provide high-level functionalities such as sets and mappings, the specification in VDM-SL tends to be more compact than the source code in Pharo.
The Pharo class for the agent definition node has 24 methods and its subclass for stage definition has 15 methods while the AST module in VDM-SL has only one function related to the agent definition, namely \ct{sizeOfAgent}.
Fields of a record value can be accessed by default as declared in the record type definition.
In object-oriented programming, those accessor methods are considered good practice for specialization by inheritance, and Pharo's programming tools provide good UIs to operate over overridden small methods with less burden.
Besides accessor methods, Pharo's flexible object system allows the programmer to define control structures as methods.
The \ct{attributeDeclarationsDo:} method of the agent definition class iterates over the attribute declarations to evaluate the closure.
In the VDM specification, those enumerations are done by the language, {\it i.e.} the sequence-binds and for-in-do statements.
%A few functions like the \ct{sizeOfAgent} function are needed to avoid repetitions.

The cost of the specification phase has been a small fraction in the development of \remobidyc.
Case studies of lightweight formal methods indicate that the formal specification improves the quality and productivity of the implementation.
The cost of specification in the \remobidyc was insignificant when the explorative tasks were done in the preceding implementation phase.
 
%%%%%%%%%%%%%%%%%%%%%%%%%%%%%%%%%%
\section{Concluding Remarks}
\label{sec:concluding-remarks}

This paper reported the development of \remobidyc as a case of the implementation-first approach of formal specification.
The development time and amount of source text were significantly smaller than those of implementation.
In successful cases of formal specification by conventional specification-first approach, specification took longer time than implementation.
It is highly probable that the source of the cost of the specification case is more in the exploratory process than in presenting in formal specification languages.
The exploration to understand the problem domain and to communicate with domain experts is required in the development of novel software.
The developers need to pay the cost of exploration regardless of which language to use in the exploration, a specification language or a programming language.
Formal specification is applicable in either case.
The authors consider that the case of \remobidyc introduced in this paper mentions that the implementation-first approach with an agile programming language can be a good choice of lightweight formal methods.
Further study is needed to understand the effect of formal specification applied in the implementation-first approach.

%%%%%%%%%%%%%%%%%%%%%%%%%%%%%%%%%%
\section*{Acknowledgments}
The authors thank the Pharo community for technically supporting the implementation platform and the anonymous reviewers for their valuable comments and suggestions.

\bibliographystyle{splncs03}

\bibliography{article}

\end{document}